\newcommand\apj{ApJ}%    % Astrophysical Journal
\newcommand\apjl{ApJL}     % Astrophysical Journal, Letters
\newcommand\apjs{ApJS}%    % Astrophysical Journal, Supplement
\newcommand\aap{A\&A}%     % Astronomy and Astrophysics
\newcommand\mnras{MNRAS}%   % Monthly Notices of the RAS
\newcommand\solphys{SoPh}% % Solar Physics
\newcommand\ssr{SSRv}% % Space Science Reviews
\newcommand\grl{GRL}%  % Geophysics Research Letters
\newcommand\jgr{JGR}%     % Journal of Geophysics Research

\newcommand\aj{AJ}%  % THE ASTRONOMICAL JOURNAL 
\documentclass[universe,article,accept,moreauthor]{Definitions/mdpi}

\firstpage{1}
\pubvolume{1}
\issuenum{1}
\articlenumber{0}
\pubyear{2026}
\copyrightyear{2026}
\datereceived{13 January 2026}
\daterevised{13 February 2026} % Comment out if no revised date
\dateaccepted{xx February 2026 }
\datepublished{xx February 2026}
\Title{Stereoscopic Observation of Recurrent Streamer Waves Driven by Successive Slow Coronal Mass Ejections}

\newcommand{\speed}[1]{#1 km~s${}^{-1}$}
\newcommand{\accel}[1]{#1 m~s${}^{-2}$}
\newcommand{\nfig}[1]{Figure~\ref{#1}}
\newcommand{\rsun}[1]{${#1}\,R_\odot$}
\newcommand{\tbl}[1]{Table~\ref{#1}}
\newcommand{\arcsec}{''}
\usepackage{marginnote}
\usepackage{marginfix}

\Author{Yuandeng Shen $^{1, *}$\orcidA{} and Reetika Tiwari $^{1}$}%, Firstname Lastname $^{2}$ and Firstname Lastname $^{2,}$*}

\AuthorNames{Yuandeng Shen and Reetika Tiwari}
\address[1]{State Key Laboratory of Solar Activity and Space Weather, School of Aerospace, Harbin Institute of Technology, Shenzhen 518055, People's Republic of China}

\corres{Correspondence: ydshen@hit.edu.cnn}
\abstract{We report the stereoscopic observations of two recurrent streamer waves in a single streamer structure, utilizing coordinated observations from the SOHO, STEREO, and SDO missions. Contrary to the long-held view that fast coronal mass ejections (CMEs) are necessary drivers, we demonstrate that these recurrent waves were excited by two consecutive slow CMEs ($<$\speed{500} accompanied by only modest flare activity. Three-dimensional reconstruction reveals that the first and second waves propagated with significant decelerations of \accel{\textminus 7.93} and \accel{\textminus 10.26}, respectively. Their average amplitudes were \rsun{0.41} and \rsun{0.77}, wavelengths were \rsun{4.02} and \rsun{6.17}, and periods were 2.66 and 2.53 hours, respectively. While the amplitude of the first wave declined with heliocentric distance (consistent with conventional energy convection), the second wave exhibited an intriguing increasing trend in amplitude. Both waves showed a linear increase in wavelength and period with distance, indicating a non-stationary and dispersive medium. Crucially, despite the disparity in driver energy and wave scales, the periods and their change rates remained nearly identical for both events. This provides compelling case-specific evidence that the streamer wave period is primarily determined by the inherent eigenmodes of the streamer plasma slab rather than the specific characteristics of the trigger. We conclude that the generation of observable streamer waves is a combined consequence of the streamer's structural stability and the energy transfer efficiency of the triggering disturbance.}


\begin{document}

\section{Introduction}
The high temperature and magnetic solar corona plasma supports the propagation of various kinds of magnetohydrodynamics (MHD) waves and oscillation modes, due to the elastic and compressible properties of the coronal plasma medium \cite{naka05, warmuth15, 2022SoPh..297...20S}. In recent decades, various wave modes have been discovered in the solar corona by using high-resolution observations. For example, the global extreme-ultraviolet (EUV) waves \cite{1998GeoRL..25.2465T, 2012ApJ...752L..23S,2012ApJ...754....7S,2013ApJ...773L..33S,2018ApJ...860L...8S,2018ApJ...861..105S,2021SoPh..296..169Z,2025ApJ...987L...3Z}, the slow magnetosonic waves in coronal plumes \cite{1997ApJ...491L.111O,deforest98}, the quasi-periodic fast-propagating magnetosonic waves \cite{liuw11,liuw12,2012ApJ...753...53S,2013SoPh..288..585S,nistico13,2013A&A...554A.144Y,2018ApJ...853....1S,2019ApJ...873...22S,2018ApJ...860...54O, 2022A&A...659A.164Z,2022ApJ...930L...5Z,2024NatCo..15.3281Z}, the sunspot waves \cite{su16a, su16b}, and other kinds of oscillations such as coronal loops \cite{berghmans99, aschwanden99,nakariakov99}, and filaments \cite{2014ApJ...795..130S,2014ApJ...786..151S,2015ApJ...814L..17S, bi14,zhang17,2023MNRAS.520.3080T,2021ApJ...923...74D,2023ApJ...959...71D, 2025Univ...11..401G}. Especially, the interaction between MHD waves and coronal structures has garnered significant attention in recent years, due to its potential application in coronal seismology and the diagnosis of wave properties \cite{2013ApJ...773L..33S, 2018MNRAS.480L..63S, 2019ApJ...873...22S, 2022A&A...659A.164Z,2024NatCo..15.3281Z, 2018ApJ...860...54O}. Recent advances and remaining challenges in both observing and modeling these diverse MHD modes have been comprehensively reviewed by \citet{2024RSPSA.48030950Z} and \citet{2022SoPh..297...20S}, highlighting their roles in coronal heating and plasma diagnostics. However, all these waves or oscillations are mainly observed in the lower corona; observations of wave phenomena in the outer corona are still very scarce \cite{1997ApJ...491L.111O,2022A&A...665A..51S}.

Helmet streamers are radially oriented, long-lived magnetic structures, characterized by an arcade base in the lower corona and a prolonged, high-density plasma sheet, often referred to as the streamer stalk, extending from the top of the arcade base to the outer corona. The physical and geometric properties of these structures have been established as essential channels for the solar wind \citep{1992SSRv...61..393K, 1997ApJ...485..875W}. Beyond their quasi-static appearance, streamers are inherently dynamic, characterized by the continuous release of plasma 'blobs' that contribute to the variability of the slow solar wind \citep{2000GeoRL..27..149W, 2007ApJ...655.1142S}. These structures are frequently impacted by large-scale disturbances such as filament eruptions and coronal mass ejections (CMEs). Previous observational studies have provided evidence of these interactions. For example, streamers are observed to be deflected by laterally expanding CMEs \cite{sime87,sheeley00}. Beyond simple deflection, streamers and pseudostreamers significantly influence the non-radial propagation and trajectory of CMEs by acting as magnetic guides or barriers \cite{2022FrASS...9.0183W, 2025ApJS..277...40L}. \citet{van02} observed splitting fine structures in a type-II radio burst, interpreting this as interaction between a CME and a remote streamer. Recent high spatial and high temporal resolution observations have revealed that the deflection of streamers is primarily caused by the shocks ahead of CMEs \cite{liu09a,kwon13}.

The response of streamers to CME interactions extends beyond mere deflections. \citet{liu09b} reported a permanent displacement of a streamer following CME interaction and attributed this to magnetic reconnection between ambient open magnetic fields and those newly opened by the CME. Furthermore, \citet{eselevich15} proposed the excitation of a blast shock wave resulting from the recovery motion of a deflected streamer following interaction with an impulsive CME. \citet{feng12} suggested utilizing the interaction between streamers and CMEs to diagnose the source region of CME-associated type II radio bursts. Moreover, since shocks ahead of CME bright fronts are often faintly discernible in white-light coronagraph observations, streamer response characteristics can serve as valuable proxies for detecting and diagnosing shocks in the outer corona \cite{gosling74,vourlidas03}. These interactions are also manifested in solar radio emission. For instance, the entry of a CME-driven shock into a dense streamer can cause a sudden drift rate change or a spectral 'bump' in Type II radio bursts, providing unique diagnostics for shock kinematics and the electron acceleration site \citep{2012ApJ...750..158K, 2012ApJ...753...21F}

When energetic CMEs impinge on nearby or remote streamers, they can induce sinusoidal wavelike motions in the streamer stalk. This impingement results in the deflection of the streamer. The subsequent oscillatory motion of the deflected streamer excites the outward propagating streamer wave along the streamer stalk. These phenomena were firstly reported by \citet{chen10}, and interpreted as a fast kink body mode carried by the streamer and caused by the impingement of the nearby CME. In \citet{chen10}, the period and phase speed of the streamer wave are about one hour and \speed{300--500} km/s, while the wavelength and amplitude are 2 -- 4 \rsun{} and a few tenths of \rsun{}, respectively. Using the method of coronal seismology, \citet{chen11} further estimated the Alfv\'{e}n speed and magnetic field strength surrounding the streamer stalk. They found that both the Alfv\'{e}n speed and magnetic field strength at certain distances from the center of the Sun decline with time. This suggests the recovery process of the disturbed plasma condition. \citet{feng11} further carried out a statistical analysis of streamer waves throughout Solar Cycle 23 with data taken by the Large Angle Spectrometric Coronagraph (LASCO (\cite{brueckner95})) onboard the {\it Solar and Heliospheric Observatory} (SOHO \cite{1995SoPh..162....1D}). Analyzing eight candidate events, their statistical analysis results suggested that (1) most of the streamer waves are driven by energetic CMEs characterized by high speeds and wide angular spans; (2) the interaction heights between streamers and CMEs are higher than the field of view (FOV) of C2 (say, $>$ 2 \rsun{}); (3) all front-side CMEs are accompanied by large energetic flares. This statistical study thus elucidated common properties of streamer waves and the associated CMEs. This statistical understanding was further expanded by \citet{2020ApJ...893...78D}, who analyzed 22 events during the the {\em Solar Terrestrial Relations Observatory} (STEREO \cite{kaiser08}) era and identified a clear correlation between period and wavelength, strongly supporting the hypothesis that streamer waves are eigenmodes of the streamer plasma slab. The growing database of such events has been supported by recent advancements in automated tracking methodologies \citep{2019ApJ...883..152D}. Furthermore, numerical MHD simulations have successfully reproduced these transverse motions as kink-mode oscillations in density-enhanced plasma slabs, confirming their nature as eigenmodes of the streamer structure \citep{2005A&A...442..351S, 2008ApJ...680.1532Z}.

The kink wave interpretation of streamer waves remains debated. \citet{feng13} reported a kink-like oscillation in a streamer, noting that the amplitude increases with distance and that it was not associated with CMEs. They interpreted this kink-like oscillation as resulting from Kelvin-Helmholtz instability driven by the sheared flow and magnetic field across the streamer plane. The interaction is particularly complex in pseudostreamer topologies, where incident fast-mode waves can undergo mode conversion into slow-mode waves near magnetic null points \cite{2024NatCo..15.2667K, 2025SCPMA..6859611L}. \citet{filippov14} reported that the axis of a streamer deflects impulsively and then returns gradually to its initial position following a CME interaction. This observation does not support the streamer kink wave interpretation proposed by \citet{chen10}, nor does it support the deformation of a streamer under the action of a propagating shock. They attributed this phenomenon to the motion of a large-scale magnetic flux rope associated with the CME. Specifically, the flux rope's motion away from the Sun perturbs the surrounding coronal magnetic-field lines, with these changes resembling the half-period of a wave propagating along the coronal streamer. For a comprehensive review of streamer wave research, one can refer to \citet{chen13}.

The interpretation of sinusoidal wavelike motions along streamer stalks remains an open question. Furthermore, recent high-resolution imaging from the Parker Solar Probe (PSP \cite{2016SSRv..204....7F}) has resolved fine-scale sub-structures within streamers that channel continual density fluctuations, adding a new layer of complexity to the interpretation of large-scale oscillations \cite{2023ApJ...945..116A, 2024ApJ...964..139P}. In this study, we present the first stereoscopic observational analysis of two successive streamer waves, utilizing observations from the STEREO and the SOHO. These waves were launched by the interactions of two consecutive CMEs that were associated with lower coronal filament eruptions. Unlike previous studies, we demonstrate for the first time that these streamer waves are excited by slow CMEs. Furthermore, the amplitude of the second streamer wave exhibits an increasing trend with heliocentric distance and time. The investigation of such recurrent eruptions and their induced oscillations within a common magnetic structure offers significant advantages for solar physics \cite{2002ApJ...566L.117Z, 2008ApJ...680..740D, 2011RAA....11..594S, 2015ApJ...815...71C, 2016ApJ...833..150L, 2017ApJ...845...94T, 2022ApJ...941...59Z, 2024ApJ...964..125S}. Since these successive waves probe the same plasma environment, they provide a uniquely controlled scenario for coronal seismology to more accurately diagnose the local magnetic field and plasma parameters \cite{2011ApJ...727L..43K, 2012ApJ...747...67Z, 2018ApJ...861..105S}. This approach not only enhances the reliability of the derived physical quantities but also allows for a clearer tracking of the dynamic evolution of the streamer structure itself over a short timescale. Section 2 details the instruments and observations employed. Observational results are presented in Section 3. Section 4 is dedicated to discussions concerning the excitation conditions and the observed increase in amplitude. Finally, Section 5 summarizes the conclusions and implications.

\section{Instruments and Observations}
\subsection{Multi-viewpoint Spacecraft Datasets}
\begin{figure}[H]
\centering
\includegraphics[width=0.8\textwidth]{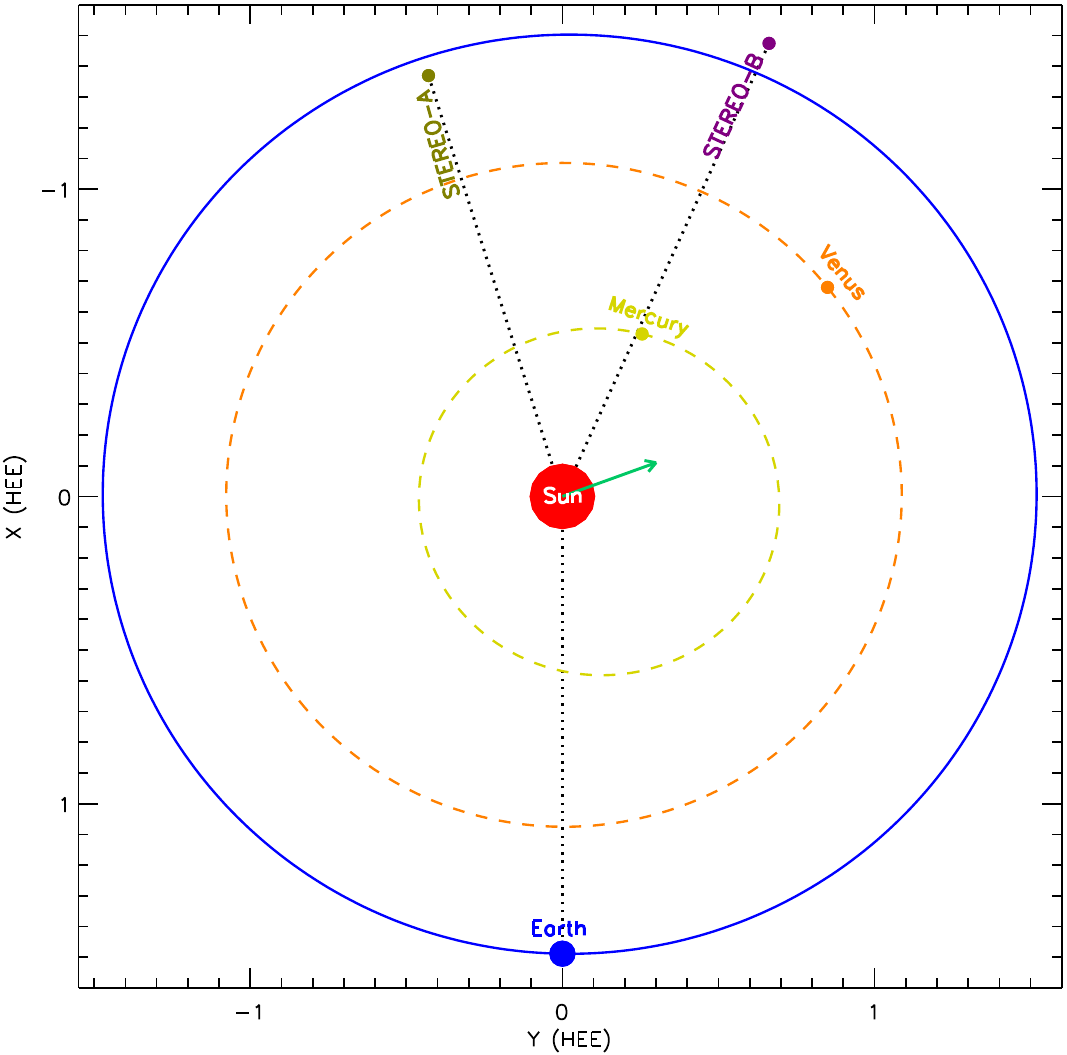}
\caption{The positions of STEREO Ahead and STEREO Behind spacecraft relative to Sun (red) and the orbit of Earth (blue) in the x-y plane of the Heliocentric Earth Ecliptic coordinate system. The green arrow points to the eruption direction of the filaments. The positions and orbits of Venus and Mercury are also plotted. The dotted lines show the angular displacement from the Sun. Units are in astronomical units (au).
\label{fig1}}
\end{figure}

To obtain a comprehensive three-dimensional (3D) perspective of the recurrent streamer waves and their driving disturbances, we combine high-resolution observations from multiple solar missions. On March 16, 2016, the separation angle between the Earth-orbiting observatories and the STEREO-Ahead (STA) spacecraft was approximately 163\textdegree. This near-quadrature configuration is particularly advantageous for stereoscopic reconstruction, as it allows for the simultaneous tracking of plasma structures from two nearly orthogonal lines of sight[cite: 154, 188]. The relative orbital positions of the relevant spacecraft and inner planets during the event are illustrated in \nfig{fig1}.

The observational data used in this study were primarily obtained from the following instrument suites:
\begin{itemize}
    \item \textbf{Low-Corona Imaging:} The eruption source regions and initial filament activities were monitored by the \textit{Atmospheric Imaging Assembly} (AIA \cite{lemen12}) onboard the \textit{Solar Dynamics Observatory} (SDO \cite{pesnell12}). We utilized the 304 \AA\ (He II) and 193 \AA\ (Fe XII/XXIV) passbands, which are sensitive to transition region and hot coronal temperatures, respectively. These images provide a pixel resolution of 0\arcsec.6 with a 12 s cadence. Complementary EUV data were provided by the \textit{Extreme Ultraviolet Imager (EUVI \cite{kaiser08})} onboard the STA spacecraft, capturing the eruptions from the off-limb perspective with a pixel size of 1\arcsec.6.
    
    \item \textbf{White-light Coronagraphy:} The propagation of the CMEs and the subsequent streamer oscillations in the extended corona were tracked using white-light coronagraphs. From the Earth's viewpoint, we utilized the \textit{Large Angle and Spectrometric Coronagraph (LASCO \cite{brueckner95})} onboard the SOHO. Data from LASCO/C2 (1.5--\rsun{6.0}) and LASCO/C3 (3.7--\rsun{30.0}) coronagraphs were employed. Simultaneously, the \textit{Sun Earth Connection Coronal and Heliospheric Investigation (SECCHI \cite{2008SSRv..136...67H})} suite onboard STA provided observations through its two coronagraphs: COR1, which monitors the inner corona from 1.4 to \rsun{4.0}, and COR2, which observes the outer corona up to \rsun{15.0}.
\end{itemize}

The technical specifications, including the FOV, spatial resolution, and temporal cadence for each instrument, are summarized in \tbl{tabinstruments} to provide a clear reference of the observational constraints. For the remainder of this paper, SOHO/LASCO/C2 and C3 are referred to as C2 and C3, respectively. Similarly, STEREO/SECCHI/COR2 onboard STEREO Ahead is denoted as COR2 and the spacecraft as STA, while SDO/AIA is referred to simply as AIA.

\begin{table}[t]
\centering
\caption{Summary of the instrumental setup, passbands, and spatio-temporal resolutions.}
\label{tabinstruments}
\begin{tabular}{lcccc}
\hline
Instrument & Wavelength / Type & FOV & Pixel Size & Cadence \\
\hline
SDO/AIA & 304 \AA, 193 \AA & Solar Disk & 0.6$''$ & 12 s  \\
STEREO/EUVI & 304 \AA, 195 \AA & Solar Disk & 1.6$''$ & 5--10 min \\
STEREO/COR1 & White Light & 1.4--\rsun{4.0} & 15$''$ & 5 min  \\
STEREO/COR2 & White Light & 2.0--\rsun{15.0} & 14.7$''$ & 15 min  \\
SOHO/LASCO/C2 & White Light & 1.5--\rsun{6.0} & 11.9$''$ & $\sim$12 min \\
SOHO/LASCO/C3 & White Light & 3.7--\rsun{30.0} & 56$''$ & $\sim$12 min \\
\hline
\end{tabular}
\end{table}

\subsection{Image Processing and Interpretation}
To clearly identify and track the wavelike motions along the streamer stalk, the running-difference technique is primarily employed. This method involves subtracting the previous intensity map from the current frame, thereby isolating moving plasma structures and suppressing the static coronal background.

Regarding the physical interpretation of these images, it should be noted that the intensity variations in running-difference maps represent the time-derivative of the emission measure rather than a direct snapshot of the absolute plasma density fluctuation ($\Delta n/n$). This approach effectively acts as a spatial-temporal filter, allowing for the precise measurement of the wave's phase velocity, wavelength, and morphological evolution. 

In this study, we define the wave crest and wave trough based on the maximum lateral displacement of the streamer axis relative to its initial equilibrium position, rather than the absolute intensity levels. In running-difference images, both the crest and trough appear as intensity enhancements (bright features) because they both represent locations where the oscillating streamer structure has moved into a previously darker background region. This geometric definition ensures a consistent tracking of the wave's phase throughout its outward propagation along the streamer stalk.

\begin{figure}[t]
\centering
\includegraphics[width=0.9\textwidth]{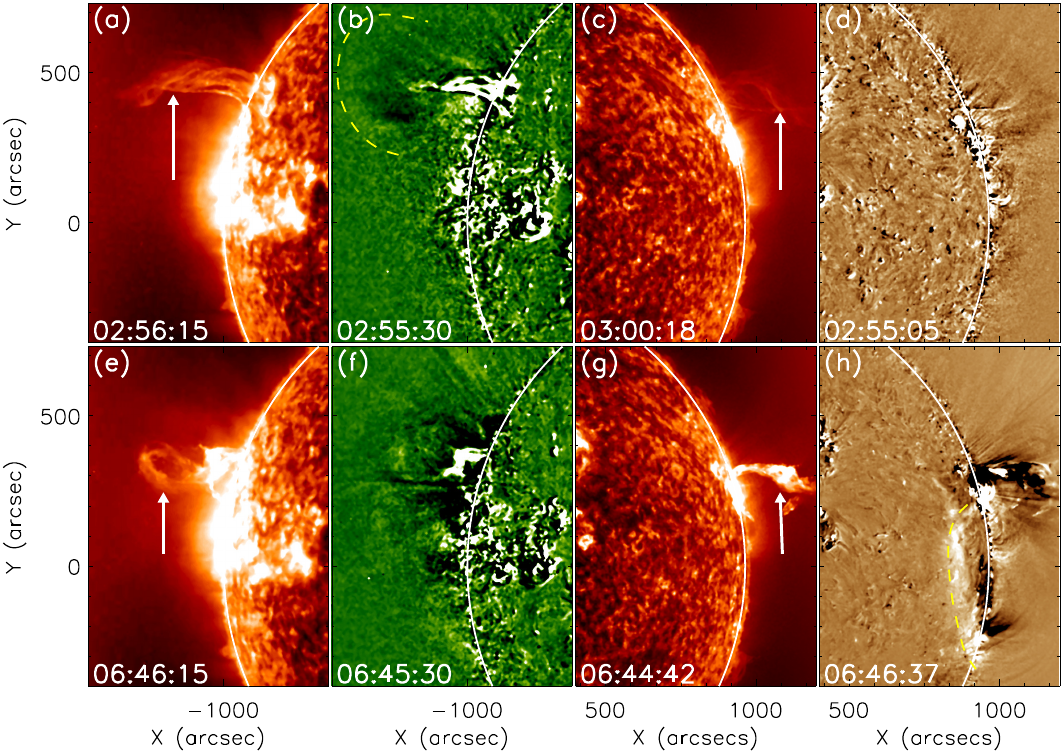}
\caption{The eruption of the two adjacent filaments. Panels (a) and (e) display STA 304 \AA\ images, while (b) and (f) show STA 195 \AA\ running difference images. Similarly, AIA 304 \AA\ images are presented in (c) and (g), and AIA 193 \AA\ running difference images in (d) and (h). White arrows in panels (a) and (c) indicate the eruption of the first filament, while those in panels (e) and (g) highlight the second filament eruption. Yellow dashed curves in panels (b) and (h) delineate the bright EUV waves associated with these two filament eruptions. The FOV for each panel is \texorpdfstring{$800\arcsec \times 1130\arcsec$}{800'' × 1130''}. An accompanying animation (animation1.mp4) is available in the online journal.
\label{fig2}}
\end{figure}

\section{Observational Results}
\subsection{Filament and CME Eruptions in the Low Corona}
On March 16, 2016, the separation angle between STA and Earth (represented by SDO and SOHO) was approximately 163\textdegree. Their observational geometry is illustrated in \nfig{fig1}, where the filament eruption direction is indicated by a green arrow. As observed by STA, the two filaments were located near the eastern limb, therefore the SDO viewed them close to the western limb. They erupted successively at approximately 02:40:00 UT and 06:35:00 UT, respectively.

The erupting filaments are depicted in \nfig{fig2}, utilizing 304 \AA\ direct images and running difference images from STA (195 \AA) and AIA (193 \AA). The eruption of the first filament is presented in the top row of \nfig{fig2}, where it is indicated by white arrows in panels (a) and (c). Within the STA 195 \AA\ running difference images, a faint, large-scale propagating EUV wave is discernible ahead of the erupting filament; its wavefront is highlighted by a yellow dashed curve in \nfig{fig2} (b). The second filament eruption is depicted in the bottom row of \nfig{fig2}. It is similarly indicated by white arrows in panels (e) and (g), and its associated EUV wave is highlighted by a yellow dashed curve in panel (h) of \nfig{fig2}. Due to the varying observational perspectives, the EUV waves accompanying the first and second filament eruptions appear less distinct in the AIA 193 \AA\ (Earth-view) and STA 195 \AA\ (STA-view) running difference images, respectively (see panels (d) and (f) in \nfig{fig2}). An animation (\texttt{animation1.mp4}) is available in the online journal, providing a more detailed view of the filament eruptions and their associated EUV waves.

\begin{figure}[H]
\centering
\includegraphics[width=0.9\textwidth]{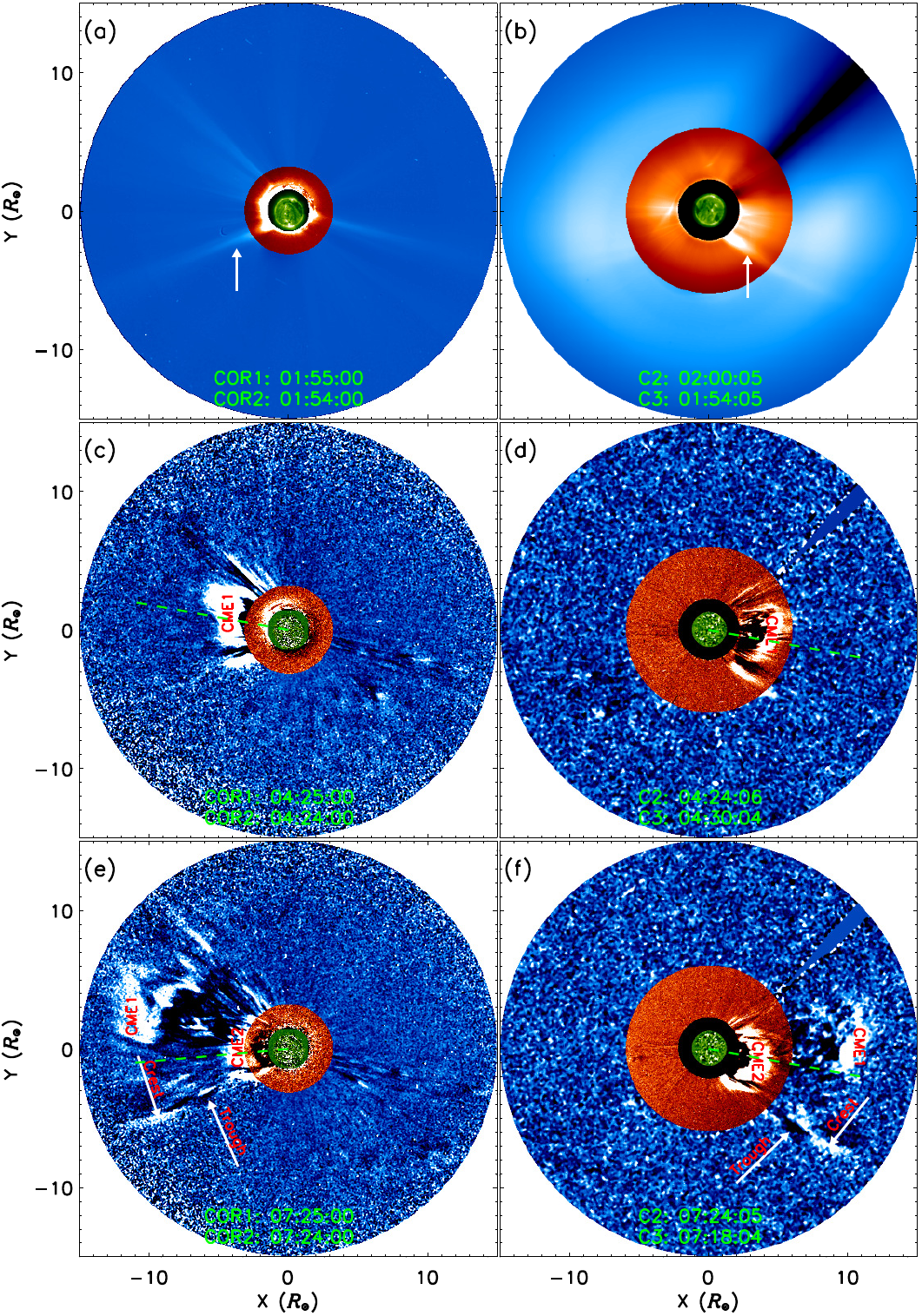}
\caption{Composite coronagraph images depict the streamer and its associated CMEs. The left and right columns display data from STEREO and SOHO-SDO, respectively. Specifically, for STEREO (SOHO-SDO) composite images, the inner, middle, and outer regions are observed by the STA 195 \AA\, (AIA 193 \AA), COR1 (C2), and COR2 (C3) instruments, respectively. Panels (a) and (b) present direct images, with two white arrows indicating the bright streamer structure. Panels (c)--(f) are running difference images. The two CMEs are designated as ``CME1'' and ``CME2'', and arrows in the bottom row indicate the wave crest and trough. Green dashed lines indicate the paths used for constructing time-distance diagrams, as illustrated in \nfig{fig4}. The approximate position angles of these green dashed lines are 285\textdegree, 105\textdegree, 277\textdegree, and 100\textdegree for panels (c)--(f), respectively. Each panel has a FOV of \rsun{30} \texorpdfstring{$\times$ \rsun{30}}. An accompanying animation (animation2.mp4) for this figure is available in the online journal.
\label{fig3}}
\end{figure}

In the outer corona, a bright, thin streamer was simultaneously observed by the SOHO/LASCO and STA (COR1 and COR2) coronagraphs. The streamer's long stalk, indicated by white arrows in \nfig{fig3}(a) and (b), was located in the southwest (position angle $\approx$ 240\textdegree) and southeast (position angle $\approx$ 120\textdegree) quadrants of the LASCO and COR2 FOVs, respectively. Two successive filament eruptions in the low corona generated two successive CMEs, which were simultaneously recorded by the C2, C3, COR1, and COR2 coronagraphs. The morphology of these two CMEs is displayed in composite running difference coronagraph images \nfig{fig3}(c) -- (f). The middle row illustrates the first CME, while the bottom row shows the second. Their first appearance of the first and the second CMEs in the LASCO C2 FOV was recorded at 03:12:09 UT and 07:00:04 UT, while in the COR2 FOV, it was at 03:24:00 UT and 07:24:00 UT, respectively. The second CME was faster than the first one, as indicated by its shorter transit time to the LASCO C2 FOV. When the second CME appeared in the C2 and COR2 FOVs, the first CME was still observable within the C3 and COR2 FOVs (see the bottom row of \nfig{fig3}). These successive CMEs impinged on the streamer, generating wavelike disturbances that propagated along its stalk. The first crest and trough of the initial streamer wave are indicated by white arrows in \nfig{fig3}(e) and (f). A detailed examination of the evolutionary processes of these two successive eruptions is provided in the associated animation (animation2.mp4), available in the online journal.

\begin{figure}[t]
\centering
\includegraphics[width=0.8\textwidth]{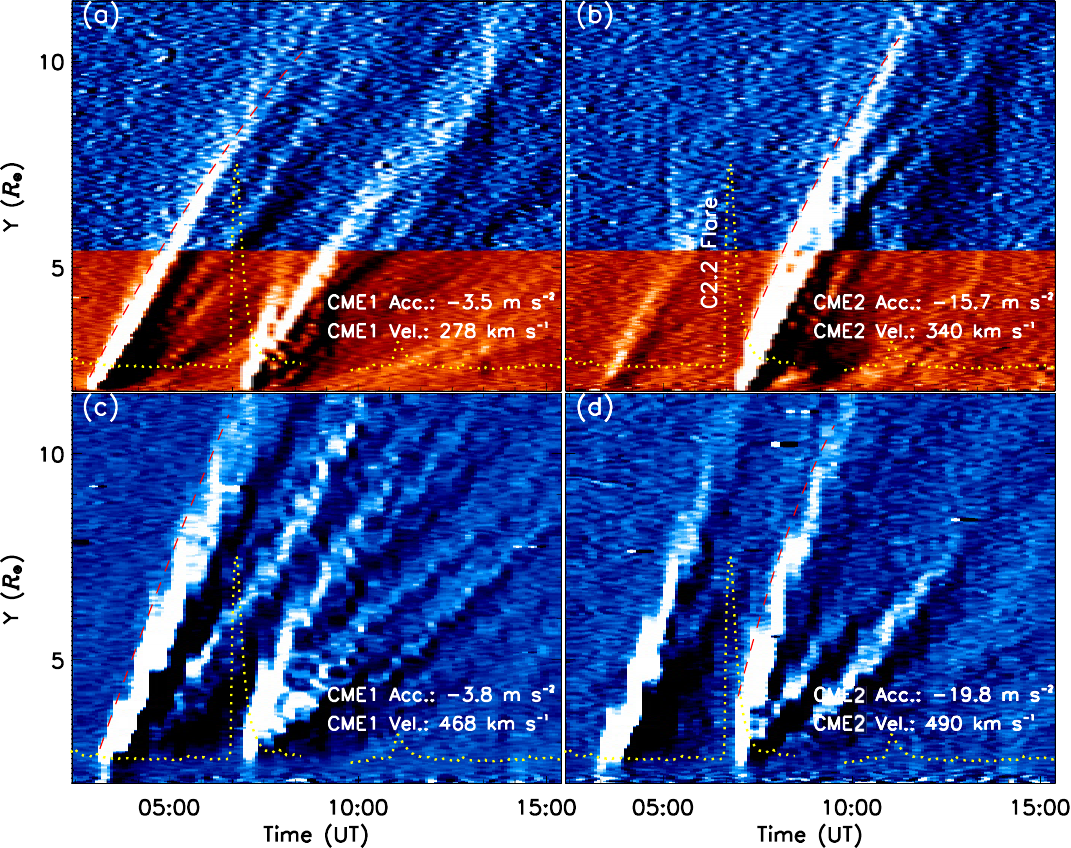}
\caption{Time-distance diagrams show the kinematics of the two successive CMEs. Panels (a) and (b) present composite time-distance diagrams, constructed from measurements along the dashed green lines indicated in \nfig{fig3} (d) and (f). Within these diagrams, the orange and green sections are derived from C2 and C3 running difference images, respectively. Panels (c) and (d) display time-distance diagrams generated from COR2 running difference images, with measurements taken along the green lines depicted in \nfig{fig3} (c) and (e), respectively. In each panel, the red dashed curve represents a second-order polynomial fit to the trajectory of the prominent emission front, while the yellow dotted curve shows the corresponding GOES soft X-ray 1 -- 8 \AA\ flux. The corresponding derived decelerations and speeds for the two CMEs are also overlaid in each panel.
\label{fig4}}
\end{figure}

Figure \ref{fig4} presents the kinematics of the two CMEs, visualized through time-distance diagrams derived from running difference coronagraph images. The paths for these diagrams are illustrated in Figure \ref{fig3} (c) and (d). In Figure \ref{fig4}, panels (a) and (b) are composite time-distance diagrams, where the orange and blue segments correspond to C2 and C3 observations, respectively. Panels (c) and (d) are constructed from COR2 running difference images. The left and right rows of Figure \ref{fig4} depict the kinematics of the first and second CMEs, respectively. The red dashed curve overlaid in each panel shows a second-order polynomial fit to the inclined bright stripe that represents the propagating CME. From LASCO observations, the speeds of the first and second CMEs are determined to be approximately \speed{278 and 340}, with corresponding deceleration of about \accel{\textminus 3.5 and \textminus 15.7}, respectively. In contrast, COR2 observations yield speeds of approximately \speed{468 and 490} and decelerations of about \accel{\textminus 3.8 and \textminus 19.8} for the first and second CMEs, respectively. These discrepancies between observation sets are likely attributable to differing view angles. 

For comparison, the online CDAW (Coordinated Data Analysis Workshops) CME catalog \footnote{\url{https://cdaw.gsfc.nasa.gov/CME_list}} reports speeds of approximately \speed{446 and 592}, decelerations of about \accel{\textminus 1.3 and \textminus 22.4}, and angular widths of 125\textdegree and 154\textdegree\ for the two CMEs, respectively. The variances between our measured results and the CDAW catalog values are probably due to variations in measuring methodologies. Collectively, these results classify both CMEs as slow CMEs, which are frequently linked to filament eruptions \cite{gosling76}. The GOES 1 -- 8 \AA\ soft X-ray flux, indicated by a yellow dotted curve, is also overlaid in each panel of \nfig{fig4}. It reveals a C2.2 flare associated with the second filament eruption, while no significant flare activity was observed for the first. This absence of an observed flare for the first event is likely because its source region was located slightly on the backside of the solar disk from Earth's perspective. However, observations from the STA view angle confirmed the presence of flare-like features, such as sudden brightening, in the source region.

\subsection{Morphology and Dynamics of the Streamer Waves}
\begin{figure}[t]
\centering
\includegraphics[width=0.9\textwidth]{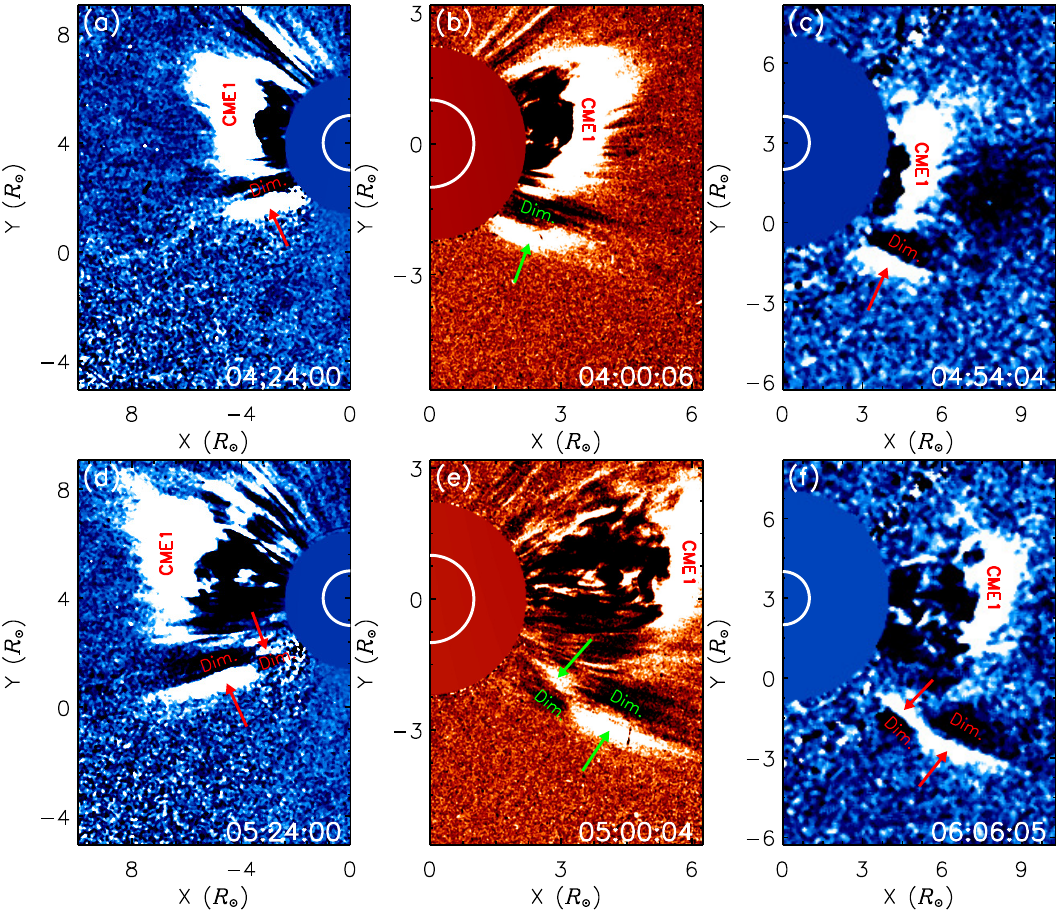}
\caption{Running difference coronagraph images displaying the first streamer wave. The left column contains COR2 running difference images, while the middle and right columns present C2 and C3 running difference images, respectively. In each panel, the inner white circle indicates the solar surface, and the shaded plate represents the occulting disk of the coronagraph. The wave crest and trough are indicated by arrows, the associated dimming regions are labeled Dim.'', and the CME is labeled CME1''. The FOVs of the three columns are \texorpdfstring{\rsun{10} × \rsun{15}, \rsun{6} × \rsun{8}, \rsun{10} × \rsun{15}}, respectively.
\label{fig5}}
\end{figure}

\nfig{fig5} illustrates the initial streamer wave observed in running difference coronagraph images from COR2 (left column), C2 (middle column), and C3 (right column). Approximately 40 minutes after the appearance of the CME in the FOVs of C2 and COR2, the lower section of the quiescent streamer stalk exhibited a distinct southward deflection motion due to impingement by the CME. In the running difference images, the deflected streamer appears as a bright, bent structure along the southern flank of the main CME, and the original equilibrium position of the streamer is marked by a dark dimming region (see the top row of \nfig{fig5}). After this impingement, the streamer subsequently exhibited pronounced largescale sinusoidal wave motions, as reported by \citet{chen10}. Due to the rebound of the deflected streamer segment, another dark dimming region appeared on the southern side of the streamer. The initial detection of the streamer wave in COR2, C2, and C3 images occurred at approximately 04:54:00 UT, 04:36:04 UT, and 05:54:04 UT, respectively. A snapshot of the streamer wave's waveform is presented in the bottom row of \nfig{fig5}. It is evident that the initial trough formed immediately behind the outward-moving crest, with the magnetic tension of the bent streamer magnetic field acting as the restoring force. Furthermore, the presence of dimming regions adjacent to the crest and trough suggests the transverse oscillation of the streamer structure. The arrows in \nfig{fig5} indicate the initial crest and trough of the streamer wave, and the dimming regions are additionally labeled 'Dim.'. 

\begin{figure}[t]
\centering
\includegraphics[width=0.8\textwidth]{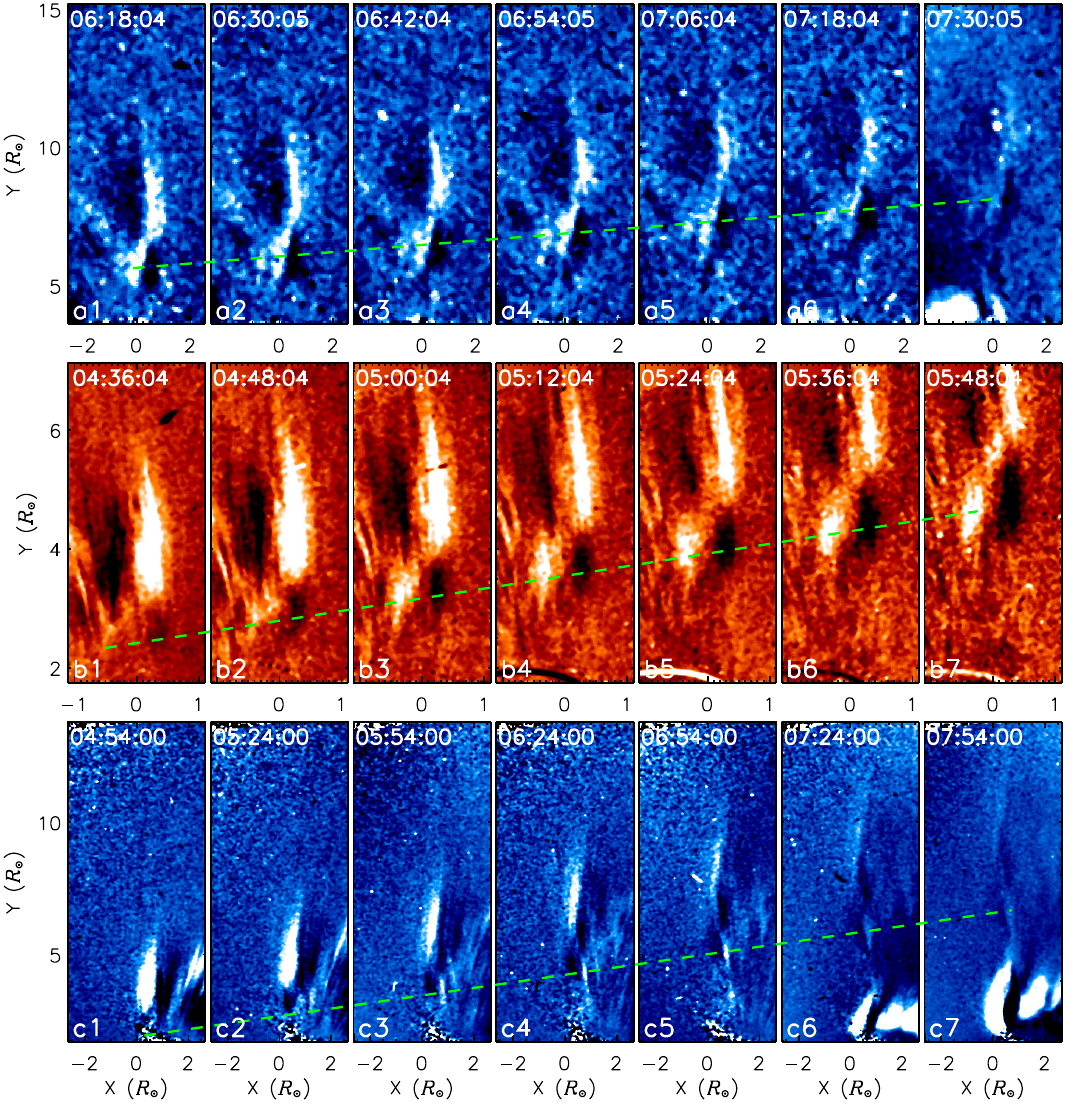}
\caption{Time sequence of running difference images showing the propagation of the first streamer wave. The first and second rows correspond to C3 and C2 running difference images, respectively, and the third row shows COR2 running difference images. The green dashed line in each row indicates the propagation of the wave trough. All images are rotated to align the streamer axis with the north. The FOVs for the three rows are \rsun{5.1} \texorpdfstring{× \rsun{11.5}, \rsun{2.2} × \rsun{5.4}, \rsun{5.3} × \rsun{12.1}}, respectively.
\label{fig6}}
\end{figure}

The detailed evolution of this initial streamer wave is depicted in \nfig{fig6}, where the top, middle, and bottom rows correspond to C3, C2, and COR2 running difference images, respectively. Notably, all images in \nfig{fig6} have been rotated so that the streamer axis points northward. In C2 and C3 images, only one complete period of the streamer wave can be visually discerned from the static sequence of running difference images. Subsequent waveforms are ambiguous and therefore not discernible in the static images; however, additional insights can be gained by viewing the accompanying animation (\texttt{animation2.mp4}) available in the online journal. The streamer wave is manifested as a sinusoidal bright structure in the running difference images, with a pair of dark dimming regions located precisely to the left (right) of the bright crest (trough). From the STA perspective, this pair of dark dimming regions is observed on the right (left) side of the crest (trough). The initial trough of the streamer wave, at different time moments, is connected by a green dashed line in \nfig{fig6}, which highlights the propagation of the streamer wave more clearly.

\begin{figure}[H]
\centering
\includegraphics[width=0.75\textwidth]{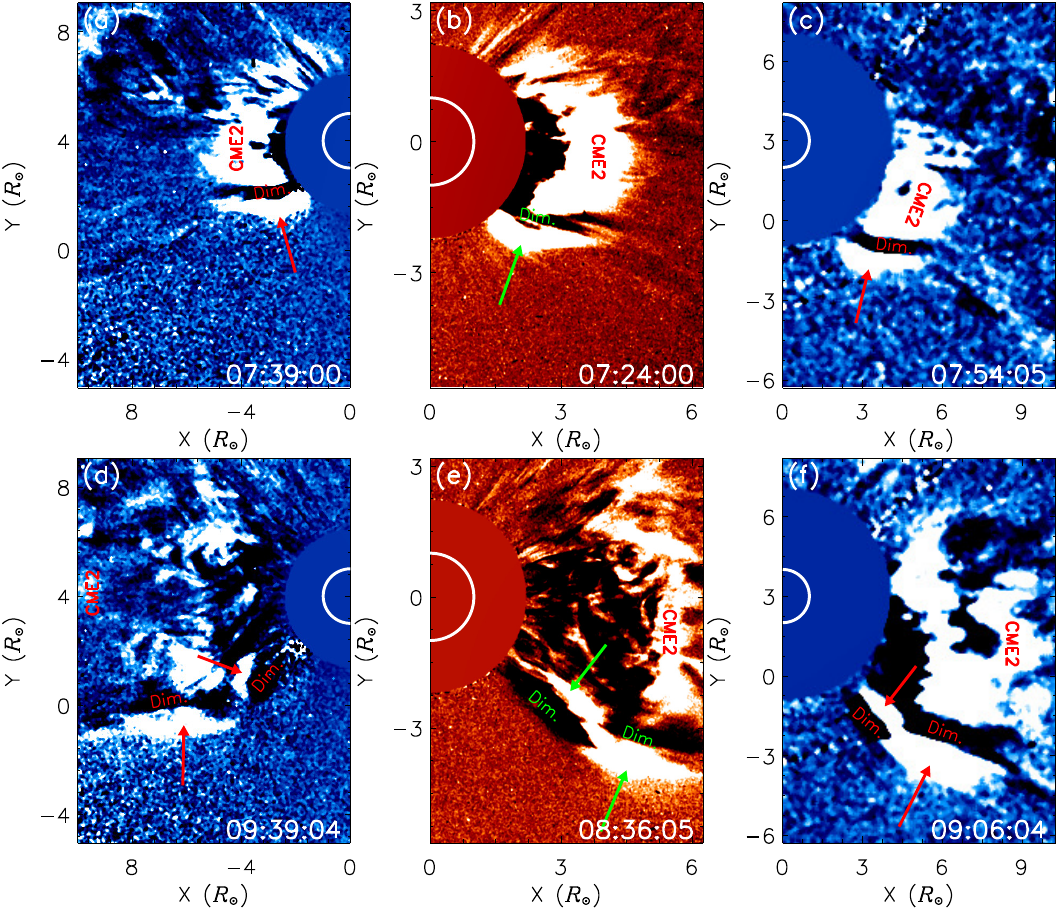}
\caption{Similar to \nfig{fig5}, but for the second streamer wave. The left column contains COR2 running difference images, while the middle and right columns present C2 and C3 running difference images, respectively. The wave crest and trough are indicated by arrows, the associated dimming regions are labeled Dim.'', and the CME is labeled CME2''. The FOVs of the three columns are \rsun{10} × \rsun{15}, \rsun{6} × \rsun{8}, \rsun{10} × \rsun{15}, respectively.
\label{fig7}}
\end{figure}

\begin{figure}[H]
\centering
\includegraphics[width=0.8\textwidth]{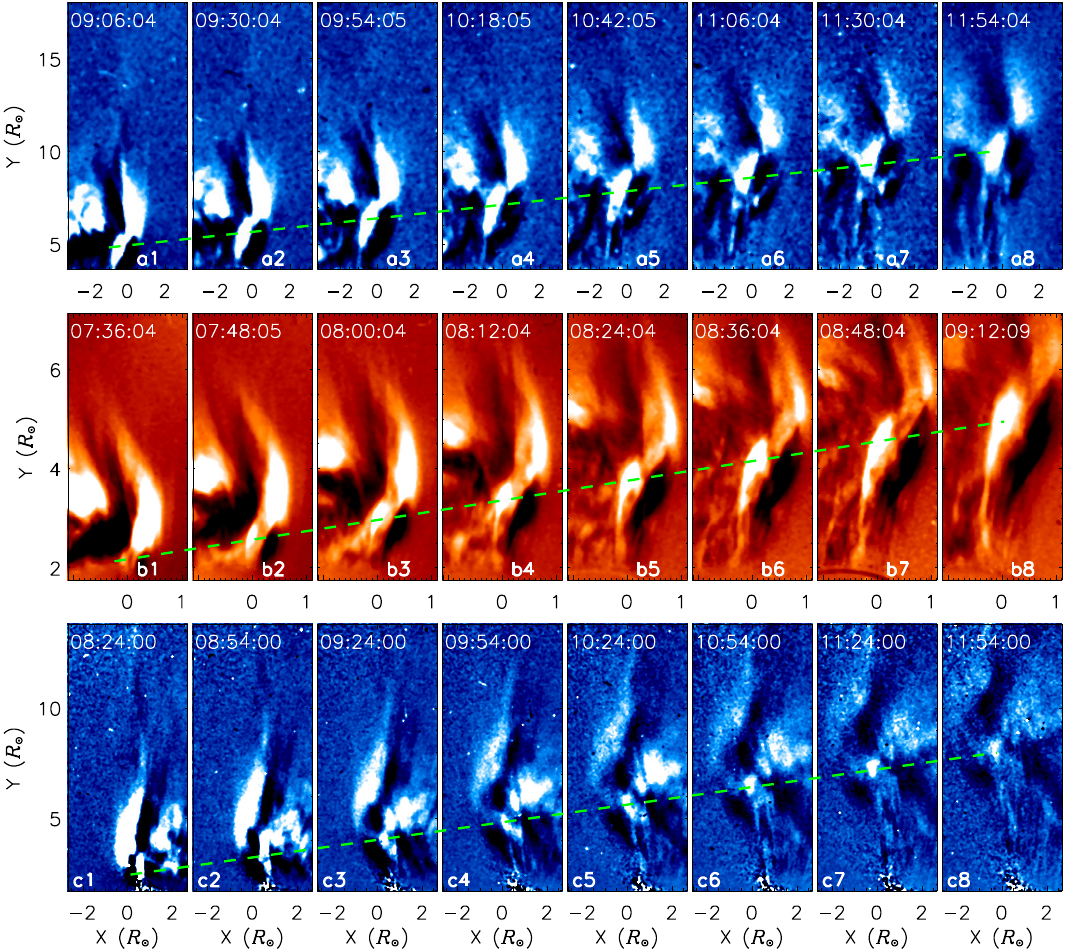}
\caption{Similar to \nfig{fig6}, but for the second streamer wave. The first and second rows correspond to C3 and C2 running difference images, respectively, and the third row shows COR2 running difference images. The green dashed line in each row indicates the propagation of the wave trough. All images are rotated to align the streamer axis with the north. The FOVs for the three rows are \rsun{6.4} × \rsun{14.4}, \rsun{2.2} × \rsun{5.4}, \rsun{5.3} × \rsun{12.1}, respectively.
\label{fig8}}
\end{figure}

\nfig{fig7} and \nfig{fig8} show the morphological structure and evolution process of the second streamer wave, respectively. Similar to \nfig{fig5}, the left, middle, and right columns of \nfig{fig7} represent COR2, C2, and C3 running difference images. The second streamer wave was triggered by the impingement of a second follow-up CME. The first appearance of the second CME in the FOVs of C2 and COR2 was at 07:00:00 UT and 07:24:00 UT, respectively, about 4 hours after the first one. It can be seen that the streamer deflection occurred immediately upon the appearance of the CME. This indicates that the interaction between the CME and the streamer occurred at a lower height, below the lower boundary of the FOVs of C2 and COR2. This is confirmed by COR1 running difference images, which show the corona from 1.4 to \rsun{4}. The morphological structure and evolution process of the second streamer wave are similar to the first one, although it exhibited a relatively larger amplitude, since the second CME was more powerful and faster than the first one. The detailed structure of the second streamer wave is shown in \nfig{fig7}. 

Similar to the first wave, the lower section of the streamer was deflected due to the impingement of the CME, and a dark dimming region appeared at the original equilibrium position of the streamer (see the top row of \nfig{fig7}). The bottom row displays the complete period of the second streamer wave. The crest and trough of the streamer wave are indicated by arrows, and the associated dark dimming regions are also marked in the figure. \nfig{fig8} shows the propagation and evolution process of the second streamer wave using a time sequence of running difference coronagraph images, in which the top, middle, and bottom rows correspond to C3, C2, and COR2 running difference images, respectively. It is observed that when the first trough of the second streamer wave appeared in the FOVs of C3 and COR2, the waveform of the first streamer wave was positioned right ahead of the first crest of the second streamer wave (see \nfig{fig8} (a1) and (c1)). The first trough of the second streamer wave at different time steps is connected by a dashed green line in \nfig{fig8}, which shows the propagation of the streamer wave more clearly. As with the first streamer wave, a pair of dark dimming regions is observed on the left (right) side of the first crest (trough) from the Earth view angle, while these are on the right (left) side of the first crest (trough) from the view angle of STA. Due to the different viewing angles, the morphology of the streamer wave appears very different in the LASCO and COR2 images.

\unskip

\subsection{Parameter Studies of the Streamer Waves}
\begin{figure}[t]
\centering
\includegraphics[width=0.8\textwidth]{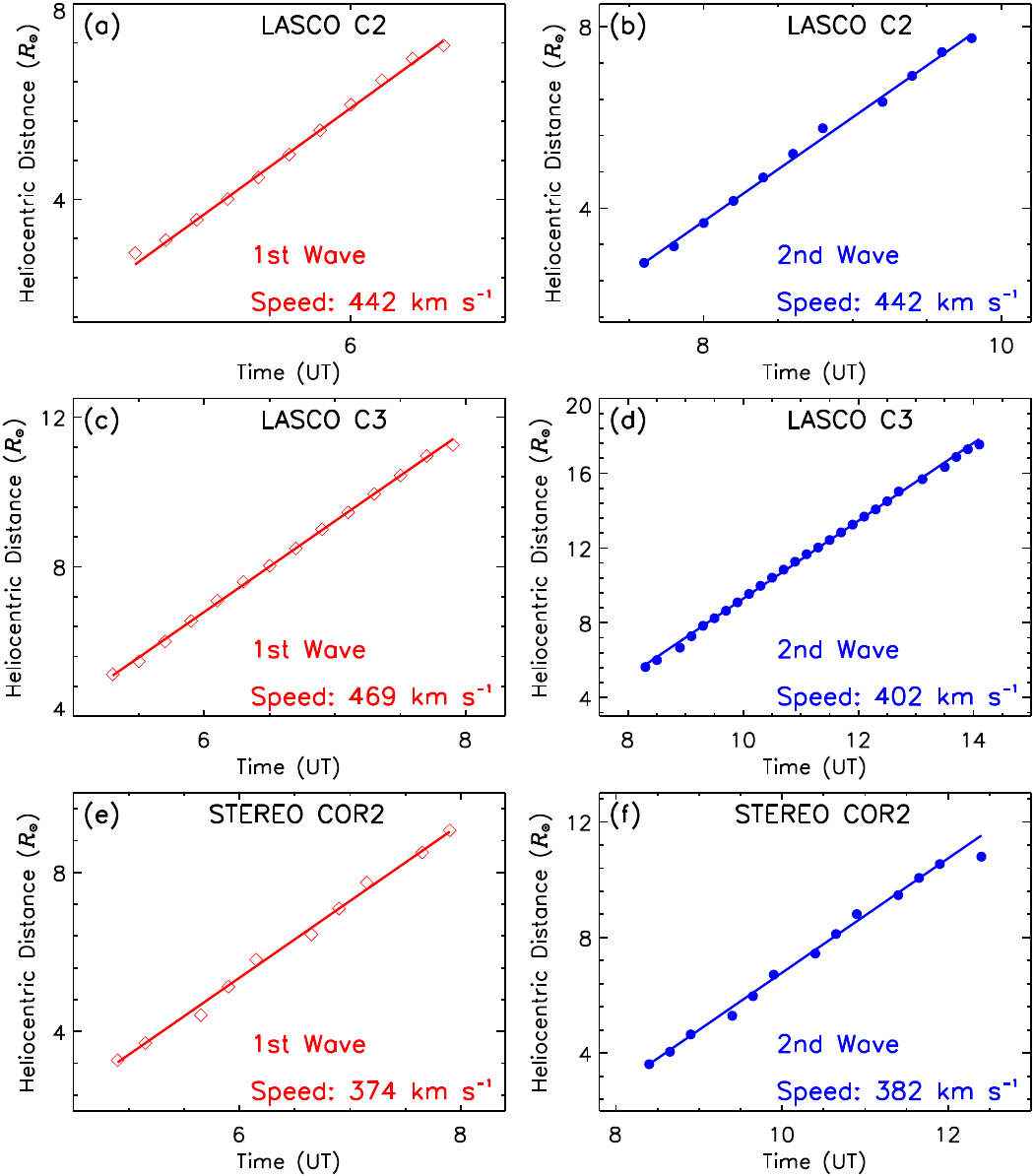}
\caption{Evolution of heliocentric distance with time for the two streamer waves. The left and right columns correspond to the first and second streamer waves, respectively. In each panel, the straight line represents the linear fit to the measured data points, with the corresponding speed plotted in the bottom right corner. The top, middle, and bottom rows display results measured from C2, C3, and COR2 observations, respectively.
\label{fig9}}
\end{figure}

Several physical parameters of the two streamer waves were measured based on C2, C3, and COR2 coronagraph observations. \nfig{fig9} shows the evolution of the heliocentric distance of the first trough as a function of time, where the left and right columns represent the results for the first and second streamer waves, respectively. The results derived from C2, C3, and COR2 images are plotted in the top, middle, and bottom rows, respectively. The results demonstrate that all measurements from instruments are well-described by a linear function. For each linear fit, the slope yields the propagation speed of the streamer wave. For the first streamer wave, the propagation speeds derived from C2, C3, and COR2 are \speed{442, 469, and 374}, while those for the second streamer wave are \speed{442, 402, and 382}, respectively. Notably, the propagation speeds of both streamer waves are consistent with the range (\speed{300 -- 500}) reported by \citet{chen10}.

\begin{figure}[t]
\centering
\includegraphics[width=0.8\textwidth]{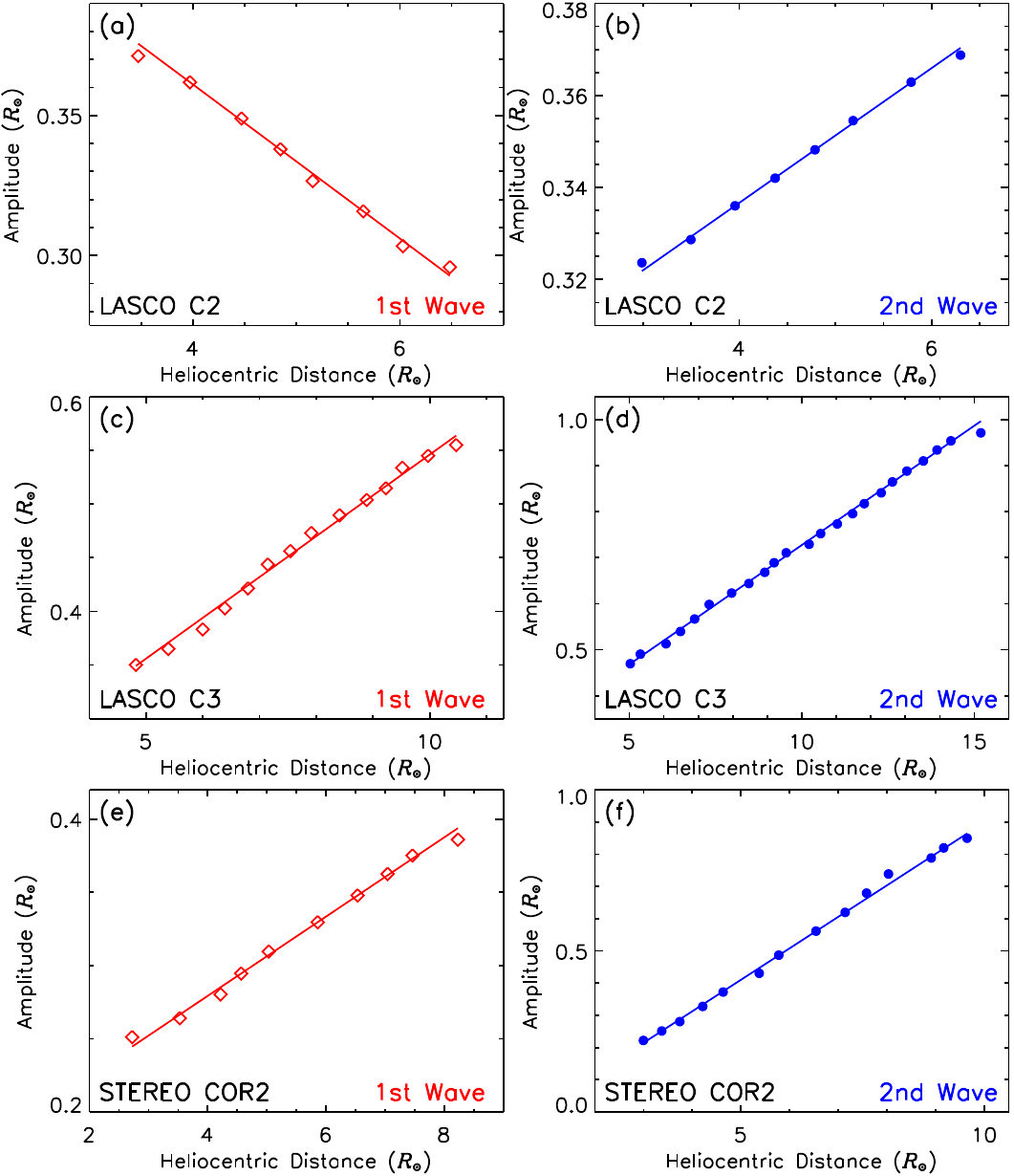}
\caption{Amplitude of the two streamer waves measured from different coronagraphs as a function of heliocentric distance. The left and right columns correspond to the first and second streamer waves, respectively. In each panel, the straight line represents the linear fit to the measured data points. The top, middle, and bottom rows display results measured from C2, C3, and COR2 observations, respectively.
\label{fig10}}
\end{figure}

The evolution of amplitude with heliocentric distance was measured for both streamer waves based on C2, C3, and COR2 observations, and the results are plotted in \nfig{fig10}. The profiles for the first and second streamer waves are shown in the left and right columns of the figure, respectively. All measured amplitudes from the different instruments exhibit a strong linear correlation with heliocentric distance. For the first streamer wave, the amplitude measured from C2 declines with increasing heliocentric distance (see \nfig{fig10} (a)), whereas those derived from C3 and COR2 increase with heliocentric distance (see \nfig{fig10} (c) and (e)). This discrepancy in trends across instruments might be attributed to the limited precision of the measurement method, given the relatively small variation range of the amplitudes. The average (median) amplitudes measured from C2, C3, and COR2 are 0.33 (0.30), 0.45 (0.46), and 0.32 \rsun{(0.32)}, respectively. 

For the second streamer wave, all three instruments (C2, C3, and COR2) consistently show an increase in amplitude with heliocentric distance (see \nfig{fig10} (b), (d), and (f)). The average (median) amplitudes measured from C2, C3, and COR2 are 0.35 (0.33), 0.72 (0.73), and 0.53 \rsun{(0.52)}, respectively. A comparison of the two events reveals that the second streamer wave possessed a larger amplitude than the first one. The larger average amplitude and wavelength of the second streamer wave are qualitatively consistent with a potentially more powerful impingement by the second CME, suggesting a more intense interaction and a larger energy transfer to the streamer structure compared to the first event. Notably, the amplitude measured from C3 images is relatively larger than those obtained from C2 and COR2, which may be due to the lower pixel resolution of C3 relative to C2 and COR2. In addition, the increasing trend observed in this study is in contrast to the results reported by \citet{chen10}, who found that the wave amplitude declines rapidly with both time and heliocentric distance.

\begin{figure}[t]
\centering
\includegraphics[width=0.8\textwidth]{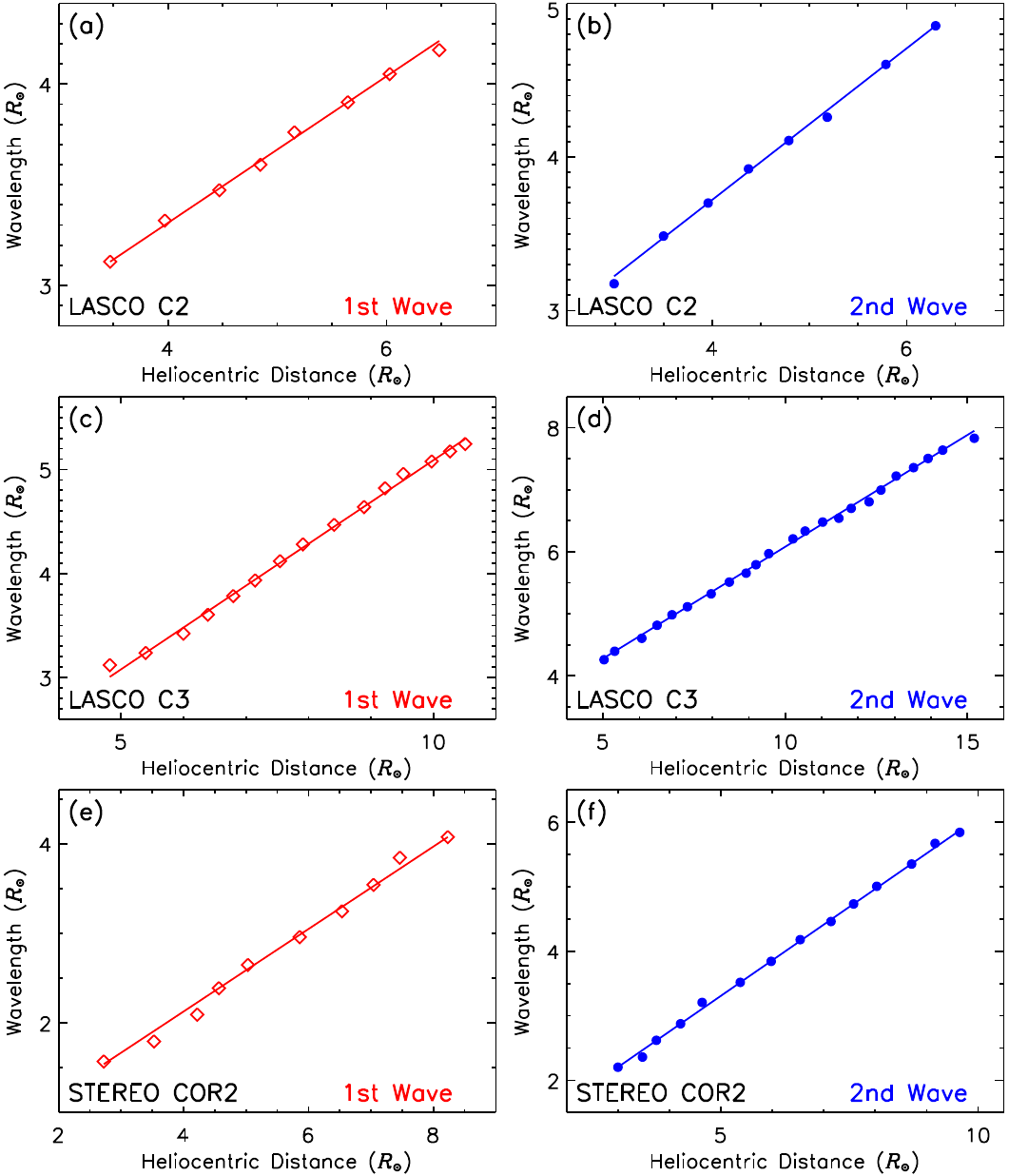}
\caption{Wavelengths of the two streamer waves measured from different coronagraphs as a function of heliocentric distance. The left and right columns correspond to the first and second streamer waves, respectively. In each panel, the straight line represents the linear fit to the measured data points. The top, middle, and bottom rows display results measured from C2, C3, and COR2 observations, respectively.
\label{fig11}}
\end{figure}

\nfig{fig11} shows the wavelength variations of the two streamer waves as a function of heliocentric distance. The left and right columns of the figure display the wavelength evolution of the first and second streamer waves, respectively, while the top, middle, and bottom rows correspond to the results derived from C2, C3, and COR2 images, respectively. The results indicate that the measured wavelengths of both streamer waves increase linearly with heliocentric distance, and can be well-fitted with a linear function. 

For the first streamer wave, the average (median) wavelengths measured from C2, C3, and COR2 are 3.67 (3.68), 4.26 (4.28), and 2.82 \rsun{(2.80)}, respectively. For the second streamer wave, the average (median) wavelengths measured from C2, C3, and COR2 are 4.01 (4.01), 6.09 (6.21), and 3.99 \rsun{(4.01)}, respectively. Consistent with the amplitude measurements, the wavelength measured from C3 images is relatively larger than those obtained from C2 and COR2, which may be attributed to the lower pixel resolution of C3 images. In addition, the wavelength values in the present case are in good agreement with the results reported by \citet{chen11}, who found that wavelength typically ranges from 2 to \rsun{4}.

\begin{figure}[h]
\centering
\includegraphics[width=0.8\textwidth]{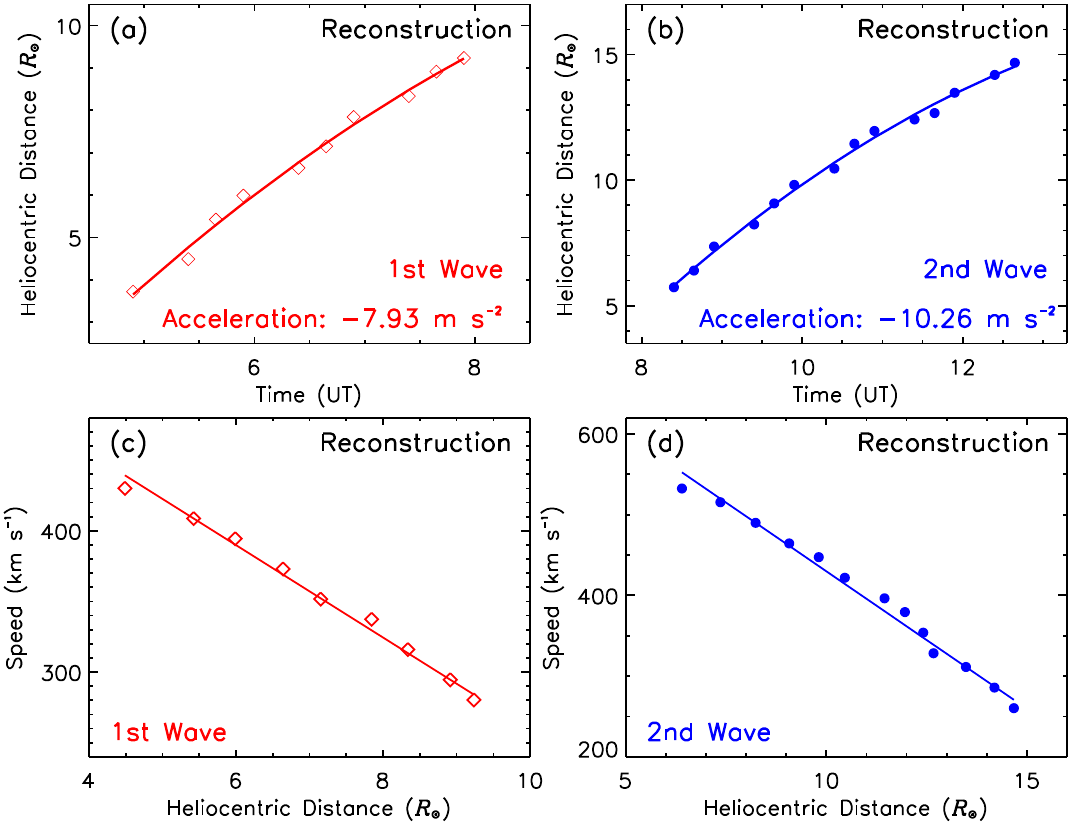}
\caption{3D reconstruction results showing the evolution of heliocentric distance with time (top row), and the derived speed profiles of the two streamer waves (bottom row). The left and right columns correspond to the first and second streamer waves, respectively. In panels (a) and (b), the curves represent a second-order fit, with the derived decelerations plotted at the bottom of each panel. In the bottom row, the straight line in each panel denotes the linear fit to the points.
\label{fig12}}
\end{figure}

\subsection{Three-Dimensional Reconstruction Results of the Streamer Waves}
The 3D structures of the two successive streamer waves were reconstructed by using multi-view coronagraph observations from LASCO and COR2 and the standard reconstruction procedure ``scc\_measure.pro'' available in the SSW package. It should be pointed out that the paired COR2 and LASCO images are not completely simultaneous due to the different cadences of the two instruments. Given that the cadences of LASCO and COR2 are 12 and 15 minutes, respectively, the maximum time difference between the paired images is 6 minutes. 

The top row of \nfig{fig12} shows the evolution of heliocentric distance with time, while the variation of propagation speed with heliocentric distance is plotted in the bottom row. In contrast to the results derived from one single view angle, the 3D reconstruction results indicate that the two streamer waves exhibit obvious deceleration during their propagation (see \nfig{fig12} (a) and (b)). Therefore, we fit the reconstructed data points with a second-order function. The results yield constant decelerations of \textminus 7.93 and \rsun{\textminus 10.26} for the first and second waves, respectively. The bottom row illustrates the propagation speed decay with heliocentric distance. One can see that the streamer waves slow down rapidly as they propagate outward. For comparison, the speeds at different heliocentric distances are listed in \tbl{tab1}. It is clear that the speed of the second streamer wave consistently exceeds that of the first one at given heliocentric distances. Furthermore, the initial speeds derived from the 3D reconstruction are larger than those measured from a single view angle. The speed of the first (second) streamer wave at \rsun{4} is \speed{455 (634)}, decreasing to \speed{128 (293)} at \rsun{14}.

\begin{table}[h]
\caption{Propagation speeds at different heliocentric distances and decelerations of the two streamer waves. \label{tab1}}
\newcolumntype{C}{>{\centering\arraybackslash}X} 
\begin{tabularx}{\textwidth}{lCCCCCCC} 
\toprule
Items & 4~\rsun & 6~\rsun & 8~\rsun & 10~\rsun & 12~\rsun & 14~\rsun & Decel. \\
 & \small (km s$^{-1}$) & \small (km s$^{-1}$) & \small (km s$^{-1}$) & \small (km s$^{-1}$) & \small (km s$^{-1}$) & \small (km s$^{-1}$) & \small (m s$^{-2}$) \\
\midrule
1st Wave & 455 & 390 & 325 & 259 & 194 & 128 & $-7.93$ \\
2nd Wave & 634 & 566 & 498 & 430 & 362 & 293 & $-10.26$ \\
\bottomrule
\end{tabularx}
\vspace{2pt}
{\footnotesize \textbf{Note:} The deceleration values are derived from the second-order fit of the 3D reconstruction results.}
\end{table}

\begin{table}[htbp]
\caption{Periods of the two streamer waves at different heliocentric distances. \label{tab2}}
\begin{tabularx}{\textwidth}{lCCCCCC} 
\toprule
Items & 4~\rsun & 6~\rsun & 8~\rsun & 10~\rsun & 12~\rsun & 14~\rsun{} \\
 & \small (hours) & \small (hours) & \small (hours) & \small (hours) & \small (hours) & \small (hours) \\
\midrule
1st Wave & 0.52 & 1.50 & 2.48 & 3.45 & 4.43 & 5.41 \\
2nd Wave & 0.76 & 1.66 & 2.55 & 3.45 & 4.35 & 5.24 \\
\bottomrule
\end{tabularx}
{\footnotesize \textbf{Note:} The periods are derived from the reconstructed wavelengths and propagation speeds ($p = \lambda/v$).}
\end{table}

\begin{figure}[t]
\centering
\includegraphics[width=0.8\textwidth]{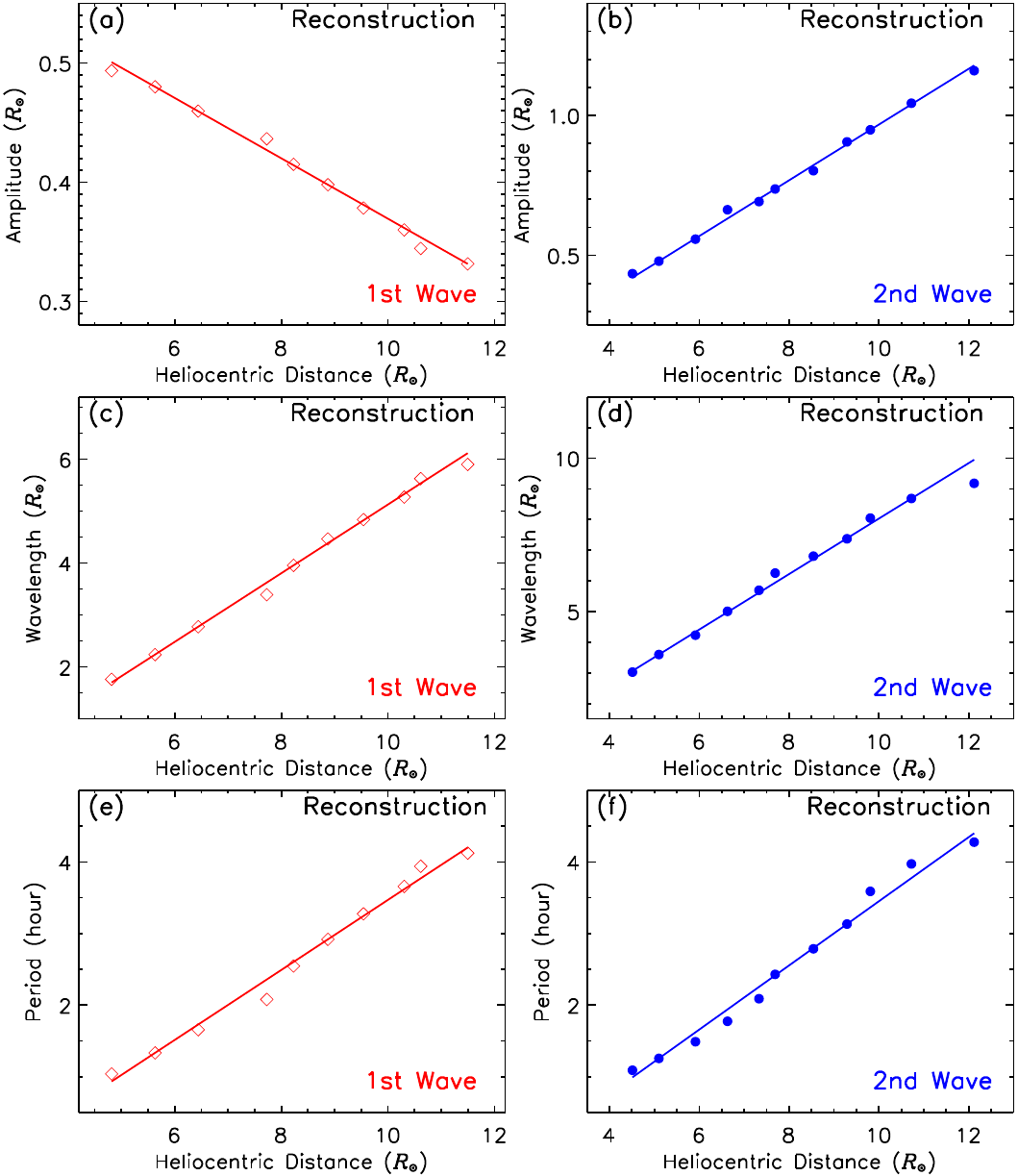}
\caption{3D reconstruction results of amplitudes (top row), wavelengths (middle row), and periods (bottom row) as a function of heliocentric distance for the two streamer waves. The left and right columns correspond to the first and second streamer waves, respectively. In each panel, the straight line represents the linear fit to the measured points.
\label{fig13}}
\end{figure}

The amplitudes, wavelengths, and periods of the two streamer waves were also obtained based on the 3D reconstruction results, and their variations with heliocentric distance are plotted in \nfig{fig13}. All parameters show a good linear correlation with heliocentric distance. The top row of \nfig{fig13} shows the amplitude evolution. The results indicate that the amplitude of the first streamer wave declines with increasing heliocentric distance, whereas that of the second streamer wave shows the opposite trend. The average and median amplitudes of the first streamer wave are 0.41 and \rsun{0.41}, while those for the second one are 0.77 and \rsun{0.74}, respectively. The middle row of \nfig{fig13} shows the wavelength evolution with heliocentric distance. The wavelengths of both streamer waves increase with heliocentric distance, with average (median) values of 4.02 (4.20) and 6.17 (6.25) \rsun{} for the first and second waves, respectively. The larger average amplitude and wavelength of the second streamer wave are qualitatively consistent with a potentially more powerful impingement by the second CME on the streamer structure, suggesting a more intense interaction and a larger energy transfer compared to the first event. The declining trend of the first streamer wave's amplitude is in agreement with the result presented in \citet{chen10}. The increasing trend of the second streamer wave's amplitude remains an intriguing feature, and we will discuss its possible physical origins in the next section.

The periods were derived from the reconstructed propagation speeds and wavelengths using the relation $p = \lambda/v$. Interestingly, the periods of both streamer waves increase with heliocentric distance, yielding average (median) values of 2.66 (2.73) and 2.53 (2.43) hours for the first and second waves, respectively. In addition, the period values at specific heliocentric distances are listed in \tbl{tab2}. The period is 0.52 (0.76) hours at \rsun{4} for the first (second) streamer wave, increasing to 5.41 (5.24) hours at \rsun{14}. It is found that while the average amplitude and wavelength of the second streamer wave are larger than those of the first one, their periods are remarkably similar. In addition, the rates of period evolution show little difference, with values of 29 and 27 minutes \rsun{\textminus 1} for the two waves, respectively. The similar periods and their comparable change rates suggest that the period of a streamer wave is governed by the inherent properties of the streamer structure, which supports wave propagation, rather than by the disturbance agents. In addition, the periods observed in the present case are larger than the values obtained by \citet{chen10}, who found a period of about 1 hour. This suggests that the physical conditions of the streamer in the current case differ significantly from those reported in \citet{chen10}.

\section{Discussions}
Our findings of stable periods across recurrent events are consistent with the statistical results of \citet{2020ApJ...893...78D}, who demonstrated that streamer waves act as eigenmodes of the streamer plasma slab, with their properties primarily determined by the inherent magnetic environment. To date, it remains unclear what physical conditions are requisite to form a streamer wave. \citet{chen10} and \citet{feng11} posited that the excitation of a streamer wave requires strict physical conditions to be satisfied. The authors proposed several necessary conditions for the generation of streamer waves, including: (1) the driving CME should be fast (e.g., \texorpdfstring{> \speed{1000}}) and possess a large angular span; (2) the source region of the driving CME should lie on the flank of the closed loops comprising the streamer base rather than beneath the streamer structure; and (3) the driving CME should be accompanied by an energetic flare. According to these studies, the requirement for fast CMEs is twofold: first, a fast CME causes a stronger impingement and a larger deflection from equilibrium, which is more likely to trigger oscillations. Second, a fast CME passes through the streamer structure in a relatively short time, providing an impulsive perturbation that minimizes subsequent interference with the streamer's restoration, allowing for the formation of coherent wavelike motions. These interpretations reasonably explain why most observed streamer waves are associated with energetic events. The observation of recurrent streamer waves driven by slow CMEs ($<500 \text{ km s}^{-1}$) in this study challenges the previous criterion that energetic drivers are a prerequisite for observable oscillations. This is supported by the recent work of \citet{2026ApJ...997L..28O}, who reported large-scale Kelvin-Helmholtz instability waves and streamer oscillations driven by a slow CME at approximately $200 \text{ km s}^{-1}$ observed by the Wide-field Imager (WISPR \cite{2016SSRv..204...83V}) onboard the PSP. These findings suggest that the visibility of streamer waves depends more on the structural stability and low Alfv\'{e}n speed within the streamer stalk than on the sheer momentum of the impacting CME.

Theoretically, any disturbance impacting a streamer can launch oscillations in its magnetic structure. However, many observations indicate that CME-streamer interactions typically cause only a single deflection \cite{sheeley00}; the successful generation of observable periodic oscillations is scarce. For instance, \citet{feng11} identified only eight streamer wave events in the LASCO database throughout Solar Cycle 23. In our view, the excitation threshold for streamer waves depends not only on the characteristics of the incoming CMEs but also on the properties of the streamer structures, such as magnetic field strength, plasma density, and relative orientation. Previous studies \cite{chen10,feng11} primarily focused on the influence of the incoming CMEs. If a streamer possesses a strong magnetic field and high plasma density, exciting a wave may indeed require a fast, energetic CME. Conversely, if a streamer has a weak magnetic field and low density, a modest slow CME may suffice to excite observable oscillations. In the present event, the recurrent waves are excited by two successive slow CMEs. This suggests that the streamer in this study likely had a weak magnetic field, low plasma density, and a large angle between the streamer axis and the incoming CMEs. Notably, it is not necessary to satisfy all mentioned conditions simultaneously. The key factor is the total energy transferred from the CME to the streamer during the interaction. As pointed out by \citet{feng11}, other factors such as the viewing angle and the presence of interfering bright structures in the corona also influence the observability of these waves.

Our 3D reconstruction reveals that while the second CME was more energetic, the average periods of the two recurrent waves are remarkably similar (2.66 and 2.53 hours). This observation provides a precise, case-specific validation of the statistical trends reported by \citet{2020ApJ...893...78D}, who analyzed 22 events and concluded that streamer waves represent the eigenmodes of the streamer plasma slab. By demonstrating that the oscillation period remains stable regardless of the driver's velocity (278 vs \speed{340} in LASCO), our results unequivocally support the hypothesis that the oscillation frequency is primarily governed by the inherent physical properties of the streamer structure (e.g., magnetic field strength and plasma density) rather than the characteristics of the impulsive disturbance. For example, although the second CME in our observation was potentionaly more energetic and transferred more energy to the streamer, the period of the streamer wave was similar to the previous one caused by a weak CME. This stability further reinforces the reliability of streamer waves as a robust tool for coronal seismology. This behavior is consistent with filament oscillations triggered by remote flares and EUV waves \cite{ramsey66,2014ApJ...795..130S,2014ApJ...786..151S}.

The 3D reconstruction results also show that the wavelengths of both streamer waves increase with heliocentric distance. However, their amplitudes exhibit opposite trends: the first wave's amplitude declines while the second wave's increases. Despite measurement uncertainties in coronagraph data, these trends are considered credible. The linear increase in wavelength can be explained by the spatial evolution of the Alfv\'{e}n speed and the wave period. Although both magnetic field strength and density decrease with distance, the more rapid decline in density leads to a net increase in the Alfv\'{e}n speed. Our reconstruction shows that the wave period also increases with distance (see \tbl{tab2}), which, combined with the Alfv\'{e}n speed evolution, results in the observed linear increase in wavelength. While the propagation speed decreases due to deceleration, the wavelength continues to increase. According to the relation $\lambda = v \cdot p$, this is primarily attributed to the significant increase in wave period (from $\sim$0.6 to $>$5 hours) as the wave propagates through the streamer stalk. This suggests that the streamer structure is a non-stationary and highly dispersive medium where the WKB approximation (p=constant) is not strictly applicable. The observed linear increase in wavelength and period suggests that the streamer stalk is a highly dispersive and non-stationary medium. This complexity is likely linked to MHD mode conversion. As proposed by \citet{2016SoPh..291.3195C}, fast-mode EUV waves can convert into slow-mode waves when they encounter magnetic quasi-separatrix layers or regions where the Alfv\'{e}n and sound speeds are comparable \cite{2016ApJ...822..106C,2017ApJ...834L..15Z,2018ApJ...863..101C}. This conversion leads to the formation of stationary or slowly propagating fronts, which might explain the non-stationary variations in wave parameters as the waves move from the middle corona into the outer heliosphere \citep{2022Galax..10...58C}.

While the declining amplitude of the first wave is consistent with the energy convection reported by \citet{chen10} and \citet{feng11}, the increasing trend of the second wave reflects a more complex energy evolution. We suggest this may result from the steep spatial gradients of the background magnetic field and density, or a continuous energy supply from the CME. Additionally, the amplitude growth of Wave 2 might be influenced by the internal substructure of the streamer. Recent findings by \citet{2024A&A...682A.168S} suggest that the interaction geometry might facilitate non-linear energy transfer, which, when coupled with potential Kelvin-Helmholtz instabilities (KHI) at the CME-streamer interface \cite{2013ApJ...774..141F}, could provide a sustained energy input to counteract the expected damping from convection. Furthermore, this anomalous evolution might be associated with complex MHD mode conversion. As recently imaged by \citet{2024GeoRL..5112664Z} in pseudostreamer topologies, fast-mode waves can convert into slow-mode waves at 3D null points, providing an alternative pathway for energy transport and trapping. Such processes, where wave energy is redistributed or partially captured within the streamer structure, could explain why the second wave exhibits distinct parameter variations as it propagates from the middle corona into the outer heliosphere.

\section{Conclusions}
Using stereoscopic observations from SOHO, STEREO, and SDO, we present the first stereoscopic analysis of two recurrent streamer waves driven by successive slow CMEs on March 16, 2016. LASCO observations showed CME speeds of \speed{278 and 340} with radial decelerations of \textminus 3.5 and \accel{\textminus 15.7}, respectively. COR2 observations yielded speeds of \speed{468 and 490} with decelerations of \textminus 3.8 and \accel{\textminus 19.8}. Our results demonstrate that streamer waves can be associated with slow CMEs and small flares, indicating that fast CMEs are not a strictly necessary condition. We propose that streamer wave generation is a combined consequence of the streamer's inherent properties and the CME's characteristics. The critical factor is the total energy transferred during the interaction.

The streamer waves exhibited different morphologies in LASCO and COR2 images due to varying viewing angles. Parameters including propagation speeds, amplitudes, wavelengths, and periods were measured using both direct imaging and 3D reconstruction. While direct images suggested constant propagation speeds, 3D reconstruction revealed obvious deceleration, with decelerations of \accel{\textminus 7.93} and \accel{\textminus 10.26} for the first and second waves, respectively. The 3D speed of the first (second) wave at \rsun{4} was \speed{455 (634)}, slowing to \speed{128 (293)} at \rsun{14}. This highlights the necessity of 3D reconstruction for determining the true physical properties of streamer waves.

The 3D results show the first wave's amplitude declining while the second's increases with distance. The average 3D amplitudes were 0.41 and \rsun{0.77}, respectively. While measurement noise is inherent in coronagraph data, the observed trends remain credible. The declining trend of the first wave is consistent with energy convection. The increasing trend of the second wave may result from the background density and magnetic field gradients or continuous energy input from the CME, though further numerical studies are required for confirmation.

Wavelengths for both waves increased linearly with distance, with 3D averages of 4.02 and \rsun{6.17}. Periods also increased with distance, averaging 2.66 and 2.53 hours, with change rates of 29 and 27 minutes \rsun{\textminus 1}. Notably, while the second wave had a larger amplitude and wavelength consistent with a more powerful driving CME, the periods and their change rates were nearly identical for both waves. This strongly suggests that the oscillation period is governed by the inherent properties of the streamer structure rather than the disturbance agents.

In summary, this study demonstrates that fast CMEs and energetic flares are not necessary for exciting streamer waves. The excitation depends significantly on the inherent properties of the streamer structure. The recurrence of waves in a single streamer confirms that oscillation periods are primarily determined by the streamer itself. These findings underscore the importance of 3D reconstruction in coronal physics and suggest that further high-resolution stereoscopic observations and numerical simulations are needed to fully elucidate the driving mechanisms of streamer waves.

\acknowledgments{The authors thank the SDO, SOHO, and STEREO science teams for providing the data utilized in this study. SDO is part of NASA’s Living With a Star (LWS) program; SOHO is a collaborative project between ESA and NASA; and STEREO is the third mission in NASA's Solar Terrestrial Probes program. The authors also extend their gratitude to the anonymous referees for their numerous valuable comments and suggestions, which significantly improved the manuscript. This work was supported by the Natural Science Foundation of China (Grant Nos. 12573059, 12173083) and the Specialized Research Fund for the State Key Laboratory of Solar Activity and Space Weather.}

\conflictsofinterest{The authors declare no conflicts of interest.}
\begin{adjustwidth}{-\extralength}{0cm}

\PublishersNote{}
\end{adjustwidth}
\end{document}